\documentclass[aps,physrev,superscriptaddress,reprint]{revtex4-2}

\usepackage[utf8]{inputenc}
\usepackage[T1]{fontenc}

\usepackage[english]{babel}

\usepackage{mathtools}
\mathtoolsset{centercolon}
\usepackage{braket, tensor}
\usepackage{float}

\usepackage{anyfontsize}
\usepackage[varvw]{newtx}
\usepackage{microtype}

\usepackage[dvipsnames]{xcolor}
\usepackage[colorlinks=true, allcolors=Blue]{hyperref}
\usepackage[capitalize]{cleveref}

\usepackage{siunitx}
\usepackage{graphicx}

\let\bs\boldsymbol
\let\t\tensor
\def\dd{\mathrm{d}} 
\def\ee{\mathrm{e}} 
\def\ii{\mathrm{i}} 

\def\half{\tfrac{1}{2}}

\newcommand{\ac}[1]{\hat a_\text{#1}^\dagger}
\newcommand{\bc}[1]{\hat b_\text{#1}^\dagger}
\newcommand{\ad}[1]{\hat a_\text{#1}^{\phantom\dagger}}
\newcommand{\bd}[1]{\hat b_\text{#1}^{\phantom\dagger}}

\def\TT{\mathcal T}
\def\RR{\mathcal R}

\usepackage{orcidlink}

\newcommand{\VDSP}{University of Vienna, Faculty of Physics, Vienna Doctoral School in Physics (VDSP), Boltzmanngasse 5, 1090 Vienna, Austria}
\newcommand{\VCQ}{University of Vienna, Faculty of Physics, Vienna Center for Quantum Science and Technology (VCQ), Boltzmanngasse 5, 1090 Vienna, Austria}
\newcommand{\TURIS}{University of Vienna, Research Platform for Testing the Quantum and Gravity Interface (TURIS), Boltzmanngasse 5, 1090 Vienna, Austria}

\begin{document}

\title{Measuring Space-Time Curvature using Maximally Path-Entangled Quantum States}

\author{Thomas B. Mieling \orcidlink{0000-0002-6905-0183}}
\email{thomas.mieling@univie.ac.at}
\thanks{\newline\newline\newline Based on the article published by APS in \textit{Physical Review A}, DOI:~\href{https://doi.org/10.1103/PhysRevA.106.L031701}{10.1103/PhysRevA.106.L031701}, under the terms of the \href{https://creativecommons.org/licenses/by/4.0/}{CC BY 4.0 license}. Further distribution of this work must maintain attribution to the author(s) and the published article’s title, journal citation, and DOI.}
\affiliation{\VDSP}
\affiliation{\VCQ}
\affiliation{\TURIS}
\author{Christopher Hilweg \orcidlink{0000-0003-0213-1234}}
\affiliation{\VCQ}
\affiliation{\TURIS}
\author{Philip Walther \orcidlink{0000-0002-4964-817X}}
\affiliation{\VCQ}
\affiliation{\TURIS}

\begin{abstract}
	Experiments at the interface of quantum field theory and general relativity would greatly benefit theoretical research towards their unification.
	The gravitational aspects of quantum experiments performed so far can be explained either within Newtonian gravity or by Einstein’s equivalence principle.
	Here, we describe a way to measure components of the Riemann curvature tensor with maximally path-entangled quantum states of light. We show that the entanglement-induced increase in sensitivity also holds for gravitationally-induced phases in Mach--Zehnder interferometers. 
	As a result, the height difference between the two interferometer arms necessary to rule out flat space-time by measuring gravity gradients can be significantly reduced.
\end{abstract}

\maketitle

Quantum field theory and general relativity are the current best theories for describing the interactions observed in nature. Despite having been confirmed independently with remarkable accuracy over the last century \cite{2011PhRvA..83e2122H,2012PhRvL.109k1807A,2014LRR....17....4W}, experiments demonstrating unambiguous signatures of both theories simultaneously are still lacking. The first experiment in this direction observed a phase shift induced by the evolution of the quantum wave function of single neutrons in spatial superposition of paths at different gravitational potentials \cite{1975PhRvL..34.1472C}. However, this and all subsequent experiments using matter-wave interferometry (see Ref.~\cite{2021QS&T....6b4014T} and references therein) do not require concepts from general relativity to interpret the experimental results — Newtonian gravity remains an adequate description \cite{2020PhRvX..10b1014R,2012CQGra..29v4010Z}.

In 2011, Zych \textit{et~al.}\ proposed using massive quantum systems with internal degrees of freedom to measure the evolved proper-time along each path of an interferometer \cite{2011NatCo...2..505Z}.
According to general relativity, the clocks evolve into different states depending on the path taken, which provides which-way-information and correspondingly reduces the interferometric fringe contrast.
Similar effects are predicted to occur in single-photon interferometry experiments \cite{2012CQGra..29v4010Z}.
While the interferometric areas required to observe this effect with photons seem to be out of reach with present-day technology, measurements of gravitationally-induced phase shifts are feasible \cite{2012CQGra..29v4011R,2017EJPQT.....4....2P,2017NJPh...19c3028H}.
Compared to the massive case, gravitational phase shift experiments with photons escape a purely Newtonian description as there is no Newtonian gravitational force.
Although it is possible to reproduce the weak-field predictions of general relativity in a Newtonian language by associating an effective refractive index to the gravitational field \cite{Schneider:1992} or by assigning an effective gravitational mass to the photon \cite{2012CQGra..29v4010Z}, such descriptions are restricted to the weak-field regime and are typically not considered as fundamental.
While phase shifts of photons in a homogeneous gravitational field can be explained using the weak equivalence principle (using notions from flat space-time only), this is no longer true for inhomogeneous fields. Thus, the study of gravitational phase shifts on photons due to inhomogeneous gravitational fields offers an interesting approach to measuring the interplay between general relativity and quantum theory.

\begin{figure}[t]
	\centering
	\includegraphics[width=0.9\columnwidth]{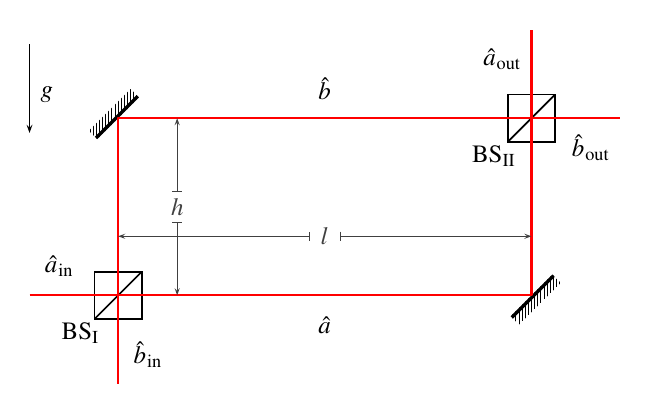}
	\caption{Schematic drawing of a Mach--Zehnder interferometer in a gravitational field. If the two input modes ($\ad{in}$, $\bd{in}$) are excited by indistinguishable photons, the first beam splitter ($\text{BS}_\text{I}$) acts as an entangling gate generating a maximally path-entangled two-photon N00N state. In the presence of gravity, the symmetry between the two arms of physical length $l$, separated vertically by a height difference $h$, is broken and the probability of obtaining both photons in either of the two output modes ($\ad{out}$, $\bd{out}$) becomes non-zero. }
	\label{fig:Mach Zehnder}
\end{figure}

In this Letter, we describe an experiment that probes the properties of maximally path-entangled optical N00N states in general static space-times. We consider a Mach--Zehnder interferometer placed vertically in Earth’s gravitational field~(\cref{fig:Mach Zehnder}). If two indistinguishable photons are sent into the two interferometer input modes, they bunch in either the upper or the lower path due to the Hong--Ou--Mandel effect \cite{1987PhRvL..59.2044H}, creating the two-photon N00N state $\tfrac{1}{\sqrt 2} (\ket{20}+\ket{02})$ \cite{2008ConPh..49..125D}. Since the phase evolution for Fock states is directly proportional to the number of photons in the mode \cite{Gerry:2004}, the two parts of the superposition above acquire the relative phase difference $2 \Delta \phi$ upon propagation through the interferometer, where $\Delta \phi$ is the phase between the arms as measured by a coherent or single-photon input state. This enhancement in phase sensitivity makes this class of states particularly useful for quantum metrology, as it allows for phase measurements at the Heisenberg limit \cite{2002JMOp...49.2325L}. In what follows, we show that the same factor-of-two enhancement in phase sensitivity over coherent and single-photon states is expected for gravitationally-induced phase shifts. The sensitivity gain provided by this maximally path-entangled quantum state can be used to measure components of the Riemann curvature tensor, which cannot be modeled using the equivalence principle alone. The extension of this formalism to N00N states with higher photon numbers is straightforward. 

To model the behavior of such multi-photon states in a static gravitational field, we consider a general static space-time, whose metric tensor, $\mathbf{g}$, can locally be written in the form
\begin{equation}
	\label{eq:metric g}
	\mathbf{g} = - c^2 N(\bs x)^2 \dd t^2 + \t h{_i_j}(\bs x)\, \dd x^i \dd x^j\,,
\end{equation}
where $c$ is the speed of light in vacuum, $N$ is the dimensionless norm of the hypersurface-orthogonal Killing vector field $\t K{^\mu} = (1, \bs 0)$ generating time translations, $\bs x = (\t x{^i})$ are local coordinates on the spatial slices of constant “Killing time” $t$, and $\t h{_i_j}$  are the components of the spatial metric tensor (i.e., the Riemannian metric induced on the hypersurfaces of constant $t$) \cite{Wald:1994}.

Neglecting the gravitational perturbation of light polarization (an effect too small to be within current experimental reach \cite{2015PhRvD..91f4041B}) the spatial modes of definite polarization decouple, and can thus be modeled by independently quantized, minimally coupled, massless scalar fields.
Restricting the discussion to a single spatial mode of definite polarization for simplicity of exposition, the electromagnetic field can thus be modeled as a quantized massless scalar field whose quantum field operator $\hat \varphi$ can be decomposed as \cite{Fulling:1989}
\begin{equation}
	\label{eq:field operator expansion}
	\hat\varphi(t, \bs x)
	= \int\negthickspace \dd \omega [
	\hat a(\omega) \ee^{-\ii \omega t} u^{\phantom\dagger}_\omega(\bs x)
	+ \hat a^\dagger(\omega) \ee^{+\ii \omega t} \bar u^{\phantom\dagger}_\omega(\bs x)
	]\,,
\end{equation}
where $\hat a(\omega)$ and $\hat a^\dagger(\omega)$ are annihilation and creation operators for photons of “Killing frequency” $\omega$ (related to proper angular frequency measured by a static observer at position $\bs x$ by $\omega_{\bs x} = \omega / N(\bs x)$), and overlines denote complex conjugation.
The mode functions $u_\omega(\bs x)$ satisfy $N^{-1} \Delta(N u_\omega) + (n \omega/c N)^2 u_\omega = 0$, where $\Delta$ is the spatial Laplacian, and $n$ is a refractive index (which is to be set to unity in vacuum).
We normalize the mode functions such that the ladder operators satisfy $[\hat a(\omega), \hat a^\dagger(\omega')] = \delta(\omega - \omega')$. In flat space-time, the mode functions can be taken to be plane waves with wave vector $|\bs k| = n\omega/c$ (and, since the discussion is restricted to a single spatial mode, such plane waves have identical propagation direction and polarization).

We assume the quantum field to be in the vacuum state which is annihilated by all $\hat a(\omega)$ operators, i.e., the ground state with respect to $\t K{^\mu}$. In the exterior Schwarzschild space-time, this corresponds to the Boulware vacuum \cite{1975PhRvD..11.1404B}.

To describe the mode coupling at a symmetric lossless beam splitter placed at $\bs x_*$, one introduces mode decompositions of the same form as in \cref{eq:field operator expansion} in each of the four input or output regions.
In flat space-time, the standard beam splitter coupling between plane wave input modes $\ad{1}, \ad{2}$ and output modes $\ad{3}, \ad{4}$, with phases chosen such that all modes have phase $-\omega t$ at $\bs x_*$, is given by
\begin{equation}
	\begin{pmatrix}
		\ad{3} \\ \ad{4}
	\end{pmatrix}
	=
	\begin{pmatrix}
		\TT	& \RR\\
		\RR	& \TT
	\end{pmatrix}
	\begin{pmatrix}
		\ad{1} \\ \ad{2}
	\end{pmatrix}\,,
\end{equation}
where $\TT$ and $\RR$ are the complex transmission and reflection coefficients of the beam splitter.
The coefficient matrix is unitary by virtue of the energy conservation relations $|\TT|^2 + |\RR|^2 = 1$ and $\bar \TT \RR + \bar \RR \TT = 0$. Since this relation applies to all modes equally (neglecting the frequency dependence of $\TT$ and $\RR$, which is admissible for sufficiently narrow spectral distributions as considered here), this implies
\begin{align}
	\hat \varphi_3(t, \bs x_*) &= \TT \hat \varphi_1(t, \bs x_*) + \RR \hat \varphi_2(t, \bs x_*)\,,\\
	\hat \varphi_4(t, \bs x_*) &= \RR \hat \varphi_1(t, \bs x_*) + \TT \hat \varphi_2(t, \bs x_*)\,,
\end{align}
which, by the equivalence principle, also holds in curved space-times.
In the case of single spatial modes considered here, this entails that in a general static space-time the beam splitter relations take the form
\begin{align}
	\begin{pmatrix}
		\ad{3}(\omega) u_{3, \omega}(\bs x_*)\\
		\ad{4}(\omega) u_{4, \omega}(\bs x_*)
	\end{pmatrix}
	=
	\begin{pmatrix}
		\TT & \RR\\
		\RR & \TT
	\end{pmatrix}
	\begin{pmatrix}
		\ad{1}(\omega) u_{1, \omega}(\bs x_*)\\
		\ad{2}(\omega) u_{2, \omega}(\bs x_*)
	\end{pmatrix}\,.
\end{align}
Note that this transformation \emph{does not} relate mode functions associated to different observers, but instead expresses the transmission and reflection of photons at a beam splitter at rest in the coordinate system considered. The appearance of the same frequencies $\omega$ on both sides of the equations is necessitated by energy conservation, and is not to be confused with the “single-mode approximation” commonly used to describe the Unruh effect \cite{2010PhRvA..82d2332B}.

For the interferometer sketched in \cref{fig:Mach Zehnder}, we choose the mode functions such that all modes at the first beam splitter and the output modes at the second beam splitter have a phase of $- \omega t$. The phases of the modes $\hat a$ and $\hat b$ in the interferometer arms, evaluated at the second beam splitter, are then fully determined by the field equations; we shall denote them by $\phi'$ and $\phi''$, respectively.
These can be computed, for example, using the geometrical optics approximation \cite{Perlick:2000}, according to which the phase shift between two monochromatic rays of equal Killing frequency $\omega$ evaluates to
\begin{equation}
	\label{eq:phase shift}
	\Delta \phi
	\equiv \omega \Delta t
	= \oint_\varGamma \frac{n \omega}{c N} \t{\hat k}{_i} \t{\dd x}{^i}\,,
\end{equation}
where $n$ is the refractive index of the medium at rest (i.e., with four-velocity equal to $c \t K{^\mu} / N$), $\t{\hat k}{_i}$ is the spatial part of the wave covector $\t k{_\mu}$, normalized with respect to the spatial metric tensor $\t h{_i_j}$, and $\varGamma = \Gamma' - \Gamma''$ is the closed loop formed by the spatial trajectory of the first ray, $\Gamma'$, followed by the second ray, $\Gamma''$, in reverse.

So far, no assumption on the gravitational field strength was made. For weak fields, however, the expressions simplify considerably: In the Newtonian \cite{Misner:1973} and also to leading-order in the post-Newtonian approximation \cite{Will:2018}, one has $N \approx 1 + \Phi/c^2$, where $\Phi$ is the gravitational potential. This leads to $\Delta \phi \approx n \omega \Delta l / c - \oint_\varGamma (n \omega \Phi / c^3) \t{\hat k}{_i} \t{\dd x}{^i}$, where $\Delta l$ is the difference in arm length (measured using the spatial metric tensor $\t h{_i_j}$).
For equally long interferometer arms placed at elevations $z'$ and $z''$, respectively, this reduces to $\Delta \phi \approx n \omega l (\Phi(z'') - \Phi(z'))/c^3$ (neglecting the vertical segments of the ray paths, which is warranted whenever horizontal variations of $\Phi$ are negligible), in agreement with the detailed analysis of fiber modes in a post-Newtonian gravitational field in Ref.~\cite{2018CQGra..35x4001B}.
Note that the further approximation $\Delta \phi \approx - n \omega l g \Delta z /c^3$ in Earth's gravitational field, with $g$ being the local gravitational acceleration, is only applicable for height differences much smaller than Earth's radius $R_\oplus$ since the series expansion of $(R_\oplus)^{-1} - (R_\oplus + \Delta z)^{-1}$ in powers of $\Delta z/R_\oplus$ converges only for $|\Delta z| < R_\oplus$ which is satisfied in low Earth orbit, but violated at the altitude of GPS satellites.

Using the general rule relating mode operators described above, the relation between the input and output mode operators can be written as
\begin{equation}
	\begin{pmatrix}
		\ac{in}(\omega) \\
		\bc{in}(\omega)
	\end{pmatrix}
	=
	\mathbf U(\omega)
	\begin{pmatrix}
		\ac{out}(\omega) \\
		\bc{out}(\omega)
	\end{pmatrix}\,,
\end{equation}
where $\mathbf U(\omega)$ is the unitary matrix
\begin{align}
	\mathbf U(\omega)
	\equiv&
	\begin{pmatrix}
		\alpha(\omega)	& -\bar\beta(\omega) \ee^{\ii \gamma(\omega)}\\
		\beta(\omega)	& +\bar\alpha(\omega) \ee^{\ii \gamma(\omega)}
	\end{pmatrix}\nonumber\\
	=&
	\begin{pmatrix}
		\TT_\text{I}	&	\RR_\text{I}\\
		\RR_\text{I}	&	\TT_\text{I}
	\end{pmatrix}
	\begin{pmatrix}
		\ee^{-\ii \phi'(\omega)}	&	0\\
		0	&	\ee^{-\ii \phi''(\omega)}
	\end{pmatrix}
	\begin{pmatrix}
		\TT_\text{II}	&	\RR_\text{II}\\
		\RR_\text{II}	&	\TT_\text{II}
	\end{pmatrix}\,,
\end{align}
where $\TT_\text{I}$, $\RR_\text{I}$ and $\TT_\text{II}$, $\RR_\text{II}$ are the transmission and reflection coefficients, of the first and second beam splitter, respectively.
Since $\mathbf U(\omega)$ is unitary, the complex parameters $\alpha = \TT_\text{I} \TT_\text{II} \ee^{-\ii \phi'} + \RR_\text{I} \RR_\text{II} \ee^{-\ii \phi''}$ and $\beta =  \RR_\text{I} \TT_\text{II} \ee^{-\ii \phi'} + \TT_\text{I} \RR_\text{II} \ee^{-\ii \phi''}$ satisfy $|\alpha|^2 + |\beta|^2 = 1$, and $\gamma$ is a real phase parameter, related to the determinant of $\mathbf U(\omega)$ by $\ee^{\ii \gamma(\omega)} = \det \mathbf U(\omega) = \ee^{-\ii(\phi'(\omega)+ \phi''(\omega))} (\TT_\text{I}^2 - \RR_\text{I}^2) (\TT_\text{II}^2 - \RR_\text{II}^2)$.

We consider a separable two-photon input state
\begin{equation}
	\ket \Psi
	= \int \dd\omega_1\, \dd \omega_2\, \psi^{(a)}_1 \psi^{(b)}_2 \ac{in,1} \bc{in,2} \ket 0\,,
\end{equation}
where, for brevity, we write $\psi^{(a)}_1 = \psi^{(a)}(\omega_1)$, and similarly for the ladder operators.
Here, the spectral functions $\psi^{(a)}$ and $\psi^{(b)}$ are taken to be normalized according to $\int \dd\omega |\psi^{(a)}(\omega)|^2 = \int \dd\omega |\psi^{(b)}(\omega)|^2 = 1$, so that $\braket{\Psi|\Psi} = 1$.
In terms of the output mode operators (we work in the Heisenberg picture), one has
\begin{align}
	\ket \Psi
	= \int \dd\omega_1\, \dd \omega_2\,& \psi^{(a)}_1 \psi^{(b)}_2 
	[ \alpha_1 \ac{out,1} - \bar\beta_1 \ee^{\ii \gamma_1} \bc{out,1}]\nonumber\\
	&\times[ \beta_2 \ac{out,2} + \bar\alpha_2 \ee^{\ii \gamma_2} \bc{out,2}]
	\ket 0\,.
\end{align}
Assuming a flat frequency response of the detectors, the probabilities $p$, of measuring one photon at each mode, and $q$, of finding both photons in the same output, evaluate to
\begin{align}
	p
	&= \int \dd\omega_1\, \dd \omega_2\, \bigg\{
	|\psi^{(a)}_1|^2 |\psi^{(b)}_2|^2 ( |\alpha_1|^2 |\alpha_2|^2 + |\beta_1|^2 |\beta_2|^2 )\nonumber\\
	&\hspace{50pt}
	- 2 \bar \psi^{(a)}_1 \psi^{(b)}_1 \bar \psi^{(b)}_2 \psi^{(a)}_2 ( \bar \alpha_1 \beta_1 \alpha_2 \bar \beta_2 )
	\bigg\}\,,
	\\
	q
	&= \int \dd\omega_1\, \dd \omega_2\, \bigg\{
	|\psi^{(a)}_1|^2 |\psi^{(b)}_2|^2 ( |\alpha_1|^2 |\beta_2|^2 + |\beta_1|^2 |\alpha_2|^2 )
	\nonumber\\
	&\hspace{50pt} + 2 \bar \psi^{(a)}_1 \psi^{(b)}_1 \bar \psi^{(b)}_2 \psi^{(a)}_2 ( \bar \alpha_1 \beta_1 \alpha_2 \bar \beta_2 )
	\bigg\}\,.
\end{align}
The consistency condition $p + q = 1$ is readily verified from $|\alpha|^2 + |\beta|^2 = 1$, and we note that the phase $\gamma$ has canceled.

For illustration, consider identical beam splitters with $\TT = 1/\sqrt 2$, $\RR = \ii / \sqrt 2$, and set both $\psi^{(a)}$ and $\psi^{(b)}$ equal to
\begin{equation}
	\label{eq:waveform}
	\psi(\omega)
	= \frac{\ee^{\ii(\omega - \omega_0) t_0}}{(2 \pi \sigma^2)^{1/4}} \exp\left(-\frac{(\omega-\omega_0)^2}{4 \sigma^2}\right)\,.
\end{equation}
In the time domain, and when evaluated at the position of the first beam splitter, $\bs x_\text{in}$, this corresponds to a Gaussian wave packet of spectral width $\sigma$ and central frequency $\omega_0$, centered at time $t_0$:
\begin{equation}
	\tilde\psi(t,\bs x_\text{in})
	= (2 \sigma^2 / \pi)^{1/4} \exp(-\sigma^2(t-t_0)^2) \ee^{-\ii \omega_0 t}\,.
\end{equation}
One then finds
\begin{equation}%
	q = \half \left(
	1 - \cos(2 \omega_0 \Delta t) \exp(- \sigma^2 \Delta t^2)
	\right)\,,
\end{equation}
or, when expressed in terms of proper frequencies $\omega_{0\text{in}}$ and $\sigma_\text{in}$ as measured at the interferometer input
\begin{equation}
	q = \half \left(
	1 - \cos(2 N_\text{in} \omega_{0\text{in}} \Delta t) \exp(- N^2_\text{in}\sigma^2_\text{in} \Delta t^2)
	\right)\,,
\end{equation}
where $N_\text{in} = N(\bs x_\text{in})$. These probabilities are identical to those obtained in flat space-time where a phase shifter is placed in one of the interferometer arms. However, this equivalence is restricted to the detection probabilities $p, q$ only, as such a phase shifter could not mimic the gravitational redshift of the wave packets arriving at the detectors (which can be demonstrated in a setup similar to the Pound--Rebka experiment \cite{1960PhRvL...4..337P}).

For comparison, we note that if only a single photon is sent into the interferometer, into mode $\ad{in}$, say, the detection probability in the $\ad{out}$ mode is ${\half[1 - \cos(\omega_0 \Delta t) \exp(- \sigma^2 \Delta t^2 / 2)]}$.
Hence, the two-photon detection probabilities are \emph{not} obtained from the one-photon probabilities by a rescaling of the time delay $\Delta t$.

The sensitivity of the effect to space-time curvature can be assessed by relating the observable probabilities to components of the Riemann curvature tensor $\t{\mathbf R}{_\mu_\nu_\rho_\sigma}$. For the space-time metric tensor given in \cref{eq:metric g}, the only non-vanishing components are
\begin{align}
	\t{\mathbf R}{_0_i_0_j}
	&= c^2 N \t{\nabla\!}{_i} \t{\nabla\!}{_j} N\,,
	&
	\t{\mathbf R}{_i_j_k_l}
	&= \t R{_i_j_k_l}\,,
\end{align}
where $\nabla$ is the spatial covariant derivative and $\t R{_i_j_k_l}$ are the components of the spatial curvature tensor \cite{Gourgoulhon:2012}. As a consequence, the Einstein vacuum equations reduce to
\begin{align}
	\Delta N &= 0\,,
	&
	\t{\nabla\!}{_i} \t{\nabla\!}{_j} N &= N \t R{_i_j}\,,
\end{align}
where $\t R{_i_j} = \t R{^k_i_k_j}$ are the components of the spatial Ricci tensor.

This shows that experiments which are sensitive only to first derivatives of $N$ cannot measure space-time curvature.
In agreement with the weak equivalence principle, the solution $\t h{_i_j} = \t \delta{_i_j}$ and $N = 1 + g z/c^2$ (which suffices to explain measurements insensitive to gravity gradients) describes accelerated coordinates in flat Minkowski space-time \cite{Misner:1973}. It follows that experiments incompatible with a linear dependence of $N$ on $z$ directly demonstrate the presence of space-time curvature and are thus beyond a description based on the weak equivalence principle. 
In the framework of Newtonian gravity, $\t\nabla{_i} \t\nabla{_j} N$ corresponds to the tidal tensor $\t\nabla{_i} \t\nabla{_j} \Phi$, which is the relevant quantity for matter-wave measurements of space-time curvature \cite{2017PhRvL.118r3602A,2022Sci...375..226O}. Note, however, that in a non-relativistic model the Newtonian gravitational field does not couple to the Maxwell equations of electrodynamics.

\begin{figure}[t]
	\centering
	\includegraphics[width=0.9\columnwidth]{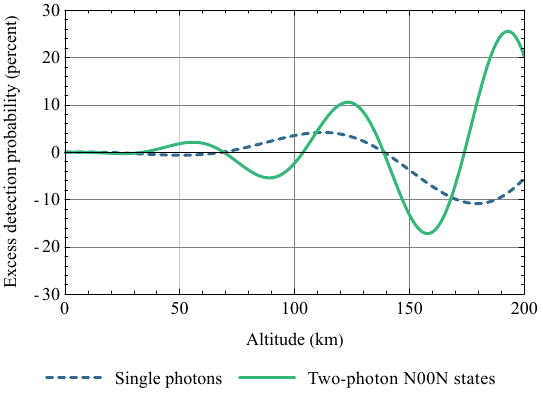}
	\caption{The difference between the detection probabilities $q$ for photons of the waveform given by \cref{eq:waveform} in a $-GM/(R_\oplus+h)$ potential and a linear potential $g h$ (excess probability) is a measure for the space-time curvature component $\t{\mathbf R}{_0_3_0_3}$. Two-photon N00N states exhibit a doubled fringe frequency and thus allow for a measurement of space-time curvature with the second arm located at lower altitudes than would be possible for coherent or single-photon states.
	For example, at height differences of $\SI{180}{\kilo\metre}$, the excess probability for two-photon N00N states is more than twice as large as for single photons, owing to the fact that the two-photon N00N-state detection probability is \emph{not} obtained from the single-photon probability by any rescaling of the time delay.}
	\label{fig:excess probability}
	\vspace*{2\baselineskip}
\end{figure}

In contrast, experiments which allow resolving second variations of the phase shift $\Delta \phi$ with height permit measurements of the space-time curvature tensor via
\begin{equation}
	\t{\mathbf R}{_0_3_0_3}
		= \frac{c^5}{n \omega l} \frac{\partial^2}{\partial h^2} \Delta \phi\,.
\end{equation}
To quantify the sensitivity to this curvature component, we plot the excess probability $\delta q = q - \tilde q$, where $q$ is computed using the $1/(R_\oplus+h)$ potential, while $\tilde q$ is computed from a linear potential (\cref{fig:excess probability}).
For illustration, we have set $l = \SI{100}{\kilo\metre}$ (which is readily accomplished using fiber optic spools), $\lambda = \SI{1500}{\nano\metre}$ and $\delta \lambda = \sigma \lambda^2/ 2 \pi c = \SI{1}{\nano\metre}$. Compared with single-photon states, the entangled N00N state exhibits double the fringe frequency and an associated increase in curvature sensitivity. In particular, N00N states allow for a significant reduction of the altitude necessary to measure space-time curvature. Measuring the second variation of the phase shift with altitude thus constitutes a direct measurement of $\t{\mathbf R}{_0_3_0_3}$. This can, for example, be realized in satellite experiments with orbits similar to those in redshift measurements by eccentric \emph{Galileo} satellites \cite{2018PhRvL.121w1101D}.

Whereas analogous effects for massive particles are explicable entirely using quantum theory in a Newtonian gravitational field, the effect described here is inherently relativistic as the equations of electrodynamics couple to gravity via the general relativistic generalization of Maxwell’s equations, while there is no generally-accepted coupling of electromagnetism to Newtonian gravity (other than taking the weak-field limit of general relativity, which is equivalent to the aforementioned methods based on either an effective refractive index of the gravitational field or an effective gravitational mass of the photon).
Moreover, due to the sensitivity to space-time curvature, the proposed experiment is not explicable using the  weak equivalence principle alone (as would be sufficient for experiments on smaller scales).
Finally, as the effect clearly relies on two-photon entanglement over distances across which the space-time curvature is non-negligible, it constitutes a genuine test for quantum field theory in curved space-time.

An experiment of this kind exhibits unambiguous signatures of both general relativity and quantum theory, as it probes the behavior of entangled two-photon N00N states in a regime of non-negligible curvature while simultaneously eluding a Newtonian description. Whereas previous proposals concerning gravitational N00N-state interferometry have suggested measurements of the gravitational redshift \cite{2021arXiv210612514A}, frame dragging \cite{2021PhRvR...3b3024B}, and parametrized post-Newtonian (PPN) parameters \cite{2021arXiv210112126R}, these analyses have disregarded gravity gradients. The analysis provided here shows that such setups can also be used to demonstrate the presence of space-time curvature at altitudes below the threshold for determining curvature with classical or single-photon probes,
and that two-photon N00N states provide a larger signal than would be produced by single photons propagating over twice the height difference.

The authors thank Piotr Chruściel, Robert Peterson and Raffaele Silvestri for useful discussions. T.M.\ is a recipient of a DOC Fellowship of the Austrian Academy of Sciences at the Faculty of Physics at the University of Vienna. P.W.\ acknowledges that this research was funded in whole, or in part, by the Austrian Science Fund (FWF) [TAI483], [F7113], [FG5], and through the research platform TURIS. 

\bibliography{bibliography}
\end{document}